\documentclass[%
superscriptaddress,
preprint,
nofootinbib,
 amsmath,amssymb,
 aps,
]{revtex4-2}

\usepackage{tabularx}
\usepackage{graphicx}
\usepackage{dcolumn}
\usepackage{bm}
\usepackage{hyperref}

\begin{document}


\title{A cloud platform for automating and sharing analysis of raw simulation data from high throughput polymer molecular dynamics simulations}

\author{Tian Xie$^*$$^\dagger$}
\affiliation{Department of Materials Science and Engineering, Massachusetts Institute of Technology, Cambridge, Massachusetts 02139, USA}

\author{Ha-Kyung Kwon$^*$$^\dagger$}
\affiliation{Toyota Research Institute, 4440 El Camino Real, Los Altos, CA 94022}

\author{Daniel Schweigert$^*$$^\dagger$}
\affiliation{Toyota Research Institute, 4440 El Camino Real, Los Altos, CA 94022}

\author{Sheng Gong}
\affiliation{Department of Materials Science and Engineering, Massachusetts Institute of Technology, Cambridge, Massachusetts 02139, USA}

\author{Arthur France-Lanord}
\affiliation{Department of Materials Science and Engineering, Massachusetts Institute of Technology, Cambridge, Massachusetts 02139, USA}
\affiliation{Sorbonne Universit\'e, Institut des Sciences du Calcul et des Donn\'ees, ISCD, F-75005 Paris, France}

\author{Arash Khajeh}
\affiliation{Toyota Research Institute, 4440 El Camino Real, Los Altos, CA 94022}

\author{Emily Crabb}
\affiliation{Department of Materials Science and Engineering, Massachusetts Institute of Technology, Cambridge, Massachusetts 02139, USA}

\author{Michael Puzon}
\affiliation{Toyota Research Institute, 4440 El Camino Real, Los Altos, CA 94022}

\author{Chris Fajardo}
\affiliation{Toyota Research Institute, 4440 El Camino Real, Los Altos, CA 94022}

\author{Will Powelson}
\affiliation{Toyota Research Institute, 4440 El Camino Real, Los Altos, CA 94022}

\author{Yang Shao-Horn$^\dagger$}
\affiliation{Department of Mechanical Engineering, Massachusetts Institute of Technology, Cambridge, Massachusetts 02139, USA}
\affiliation{Department of Materials Science and Engineering, Massachusetts Institute of Technology, Cambridge, Massachusetts 02139, USA}

\author{Jeffrey C. Grossman$^\dagger$}
\affiliation{Department of Materials Science and Engineering, Massachusetts Institute of Technology, Cambridge, Massachusetts 02139, USA}

\def\thefootnote{*}\footnotetext{These authors contributed equally to this work}
\def\thefootnote{$\dagger$}\footnotetext{Corresponding authors}



\date{\today}

\begin{abstract}

Open material databases storing hundreds of thousands of material structures and their corresponding properties have become the cornerstone of modern computational materials science. Yet, the raw outputs of the simulations, such as the trajectories from molecular dynamics simulations and charge densities from density functional theory calculations, are generally not shared due to their huge size. In this work, we describe a cloud-based platform to facilitate the sharing of raw data and enable the fast post-processing in the cloud to extract new properties defined by the user. As an initial demonstration, our database currently includes 6286 molecular dynamics trajectories for amorphous polymer electrolytes and 5.7 terabytes of data. We create a public analysis library at \url{https://github.com/TRI-AMDD/htp_md} to extract multiple properties from the raw data, using both expert designed functions and machine learning models. The analysis is run automatically with computation in the cloud, and results then populate a database that can be accessed publicly. Our platform encourages users to contribute both new trajectory data and analysis functions via public interfaces. Newly analyzed properties will be incorporated into the database. Finally, we create a front-end user interface at \url{https://www.htpmd.matr.io/} for browsing and visualization of our data. We envision the platform to be a new way of sharing raw data and new insights for the computational materials science community.

\end{abstract}

\maketitle


\section{Introduction}

In the past decade, the rapid development and application of computational theory, methodology, and infrastructure for high throughput materials discovery have generated huge amounts of data in the computational materials science community~\cite{doi:10.1146/annurev-matsci-082019-105100, LIU2017159, doi:10.1021/acs.accounts.0c00785, NANDY2022100778}. Open databases like Materials Project~\cite{doi:10.1063/1.4812323}, AFLOW~\cite{Curtarolo_2012}, and Materials Cloud~\cite{talirz2020materials} store millions of material structures and computed properties, spanning inorganic crystals, metal organic frameworks, and many other types of materials. In addition, open source software such as pymatgen~\cite{ONG2013314}, atomate~\cite{MATHEW2017140}, FireWorks~\cite{https://doi.org/10.1002/cpe.3505}, and RDKit~\cite{landrum2016rdkit} have streamlined the analysis and visualization of materials data, significantly simplifying tasks like computing effective mass from band structures~\cite{doi:10.1021/acs.chemmater.8b03529, hautier2013identification, C7TC00528H}, calculating Li-ion conductivity from molecular dynamics (MD) trajectories~\cite{SIVONXAY2020135344, doi:10.1021/acsami.0c10000, QI2021100463, D1TA06338C}, and rendering chemical structures~\cite{Flam_Shepherd_2021, molecules25112487, doi:10.1021/acs.jmedchem.7b00809}. In the biophysics community, tools such as GPCRmd~\cite{rodriguez2020gpcrmd}, BIGNASIM~\cite{10.1093/nar/gkv1301}, Cyclo-lib~\cite{10.1093/bioinformatics/btw289}, and Dynameomics~\cite{VANDERKAMP2010423} have enabled interactive analysis and visualization of MD data of proteins and small molecules. 

Despite a push toward open science, a significant portion of computational materials data has not been shared publicly~\cite{https://doi.org/10.1002/adts.201900131, ward2014making, https://doi.org/10.1002/advs.201900808} -- the raw outputs from the simulations, such as the trajectories from MD simulations and the charge densities from density functional theory (DFT) calculations. The raw data typically require gigabytes of storage for a single calculation and can easily accumulate to terabytes from a high throughput screening project. Due to the high cost of data storage and transfer, most open databases only store key properties extracted from the raw data\cite{https://doi.org/10.1002/advs.201900808, doi:10.1146/annurev-matsci-070214-020844}, while leaving the raw data in the local storage of large supercomputer centers where it is often left unattended or deleted after a period of time. Very recently, the Materials Project~\cite{doi:10.1063/1.4812323} has started to provide charge density distributions from DFT to users. However, the transfer of charge density data is not automated as users still need to communicate with the provider for access. 

\begin{figure}[tbh]
    \centering
    \includegraphics[width=\linewidth]{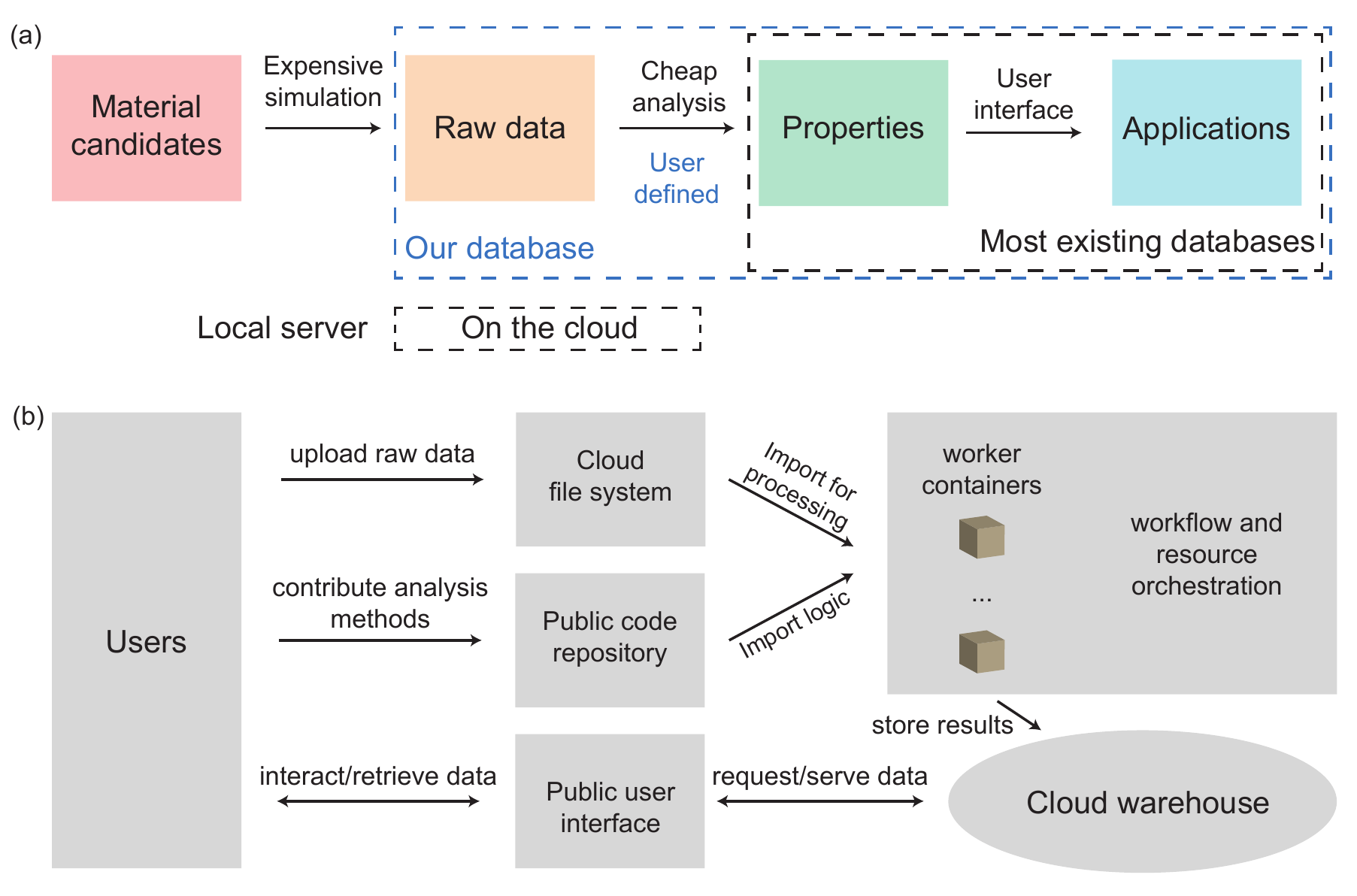}
    \caption{Illustration of the cloud based database for raw data sharing. (a) A typical workflow for computational material data generation. Our database enables the sharing of raw simulation data and flexible user-defined analysis methods to obtain properties, while most existing databases have a fixed set of pre-computed properties. (b) An overview of the software components in the platform. The platform consists of a cloud file system containing raw data, a public code repository for analysis functions, and a user interface for interacting with the data.}
    \label{fig:illustrative}
\end{figure}

In this work, we aim to provide a cloud-based platform to facilitate the sharing of raw data from high throughput material screening. Our platform includes three components, as illustrated in Figure \ref{fig:illustrative}: 1) cloud storage on Amazon Web Services (AWS) that stores raw data from simulations; 2) an open codebase on github that analyzes raw data and extracts key properties; 3) a graphical interface that allows users to interact with and visualize analyzed properties. Users can access extracted properties like in other open databases, and they can also develop new analysis functions to extract new properties from the raw data via the open codebase. Our platform eliminates the high cost of transferring terabytes of raw data by running the analysis in the cloud but still allows the user to analyze raw data based on their needs. Finally, it also aims to create a standard data format and analysis software ecosystem for MD trajectories that can eventually be expanded to include other raw outputs from other simulation methods such as DFT. We demonstrate the effectiveness of this platform by creating an open database of MD trajectories for amorphous polymer electrolytes generated using the Large-scale Atomic/Molecular Massively Parallel Simulator (LAMMPS) program \cite{plimpton1995a,thompson2022a}, which includes 6286 trajectories and 5.7 terabytes of data. 

\section{Software infrastructure and database}
\begin{figure}[tbh]
    \centering
    \includegraphics[width=\linewidth]{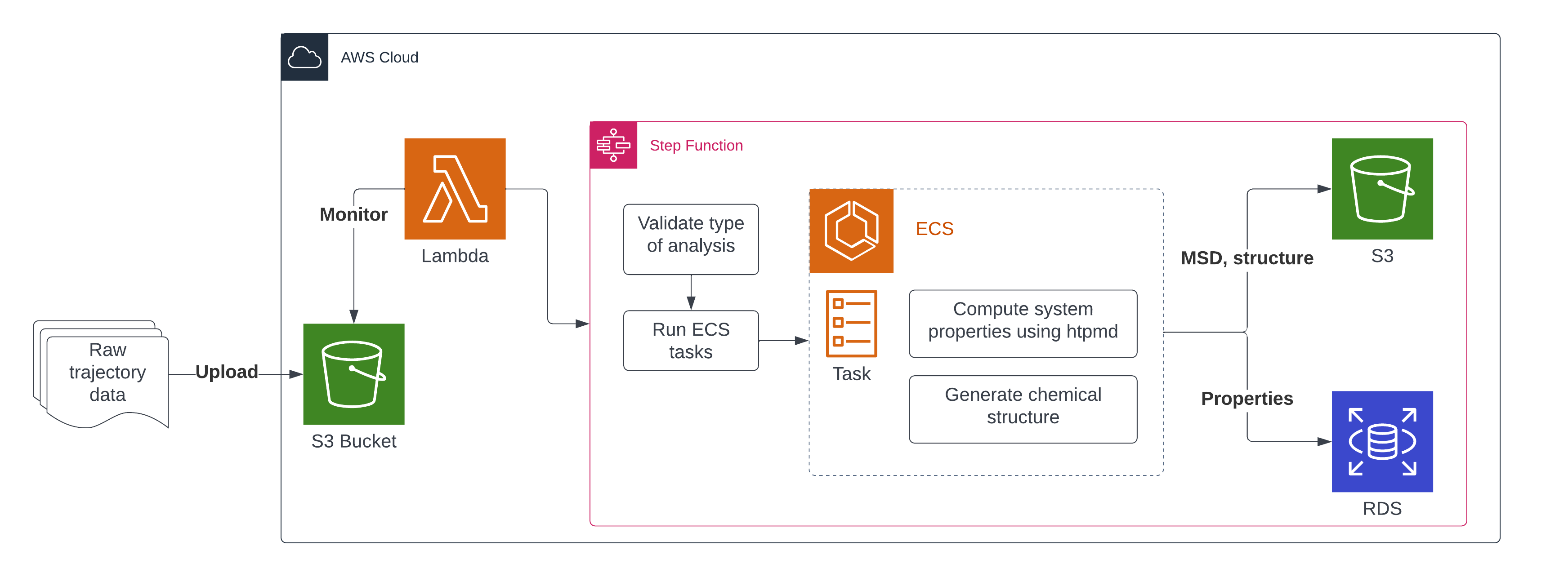}
    \caption{Detailed software workflow showing the backend processes. Final results stored in S3 bucket and RDS are accessed by the frontend UI via an API stored in a AWS Lambda instance.}
    \label{fig:software}
\end{figure}
Our software infrastructure is hosted on Amazon Web Services (AWS)~\cite{mathew2014overview} and utilizes its serverless cloud services for data processing, analysis, storage, and flow management, as shown in Figure \ref{fig:software}.

For processing, each raw trajectory data are expected to have the complete trajectory (in the "custom" LAMMPS trajectory format~\cite{lammps-dump} as well as a metadata file (in json format) that describes the input parameters of the data (SMILES, temperature, molality, length of simulations, force field, and ion types). When new trajectory data are uploaded to the platform, they are archived in an AWS Simple Storage Service (S3) bucket, which is a scalable cloud file system. The creation of the new data files in the bucket triggers an upload event, which is picked up by a serverless compute service AWS Lambda. The Lambda instance verifies the completeness of the trajectory data to ensure all files needed for the analysis are present. Afterwards, the Lambda instance initiates a workflow execution in a AWS StepFunction graph. 

As part of the StepFunction workflow, containerized AWS Elastic Container Service (ECS) tasks are run to analyze the raw trajectory data and store the results. Analyzed properties, such as ionic conductivity, diffusion coefficients of cations, anions, and polymer chains, and transference number, as well as metadata -- SMILES, molality, temperature, length of simulations, force field, and ion types -- are stored in an AWS Relational Database Services (RDS) postgreSQL database. Other types of data, such as the mean squared displacement (MSD) time series for the ions, as well as the final structure file (.cif) and an image of the monomer chemical structure (.png) are uploaded to an S3 bucket and their URLs stored in the database.

The specific analysis steps which are run as ECS tasks are specified in the public htpmd github repository (\url{https://github.com/tri-amdd/htp_md}. The repository contains analysis code suited for polymer electrolytes based on an LiTFSI salt, and allows extracting property results such as Li-ion conductivity, diffusion coefficients of Li$^+$, TFSI$^-$, and polymer chains, and transference number. In addition, the code generates average mean squared displacement (MSD) time series for Li$^+$ and TFSI$^-$ ions, as well as the final structure of the simulation box, which can also be retrieved from the UI. In addition, pre-trained machine learning models are applied to the existing trajectory data to provide predictions of a subset of the properties. More details are found in a later section. While there is a natural discrepancy between predicted and computed properties due to the variance in the machine learning models, predictions can be useful for making estimates of a property without having to run the full length of the simulation.

Members of the research community are encouraged to provide their specific analyses or prediction models to the github codebase via a pull request from a fork. Upon review and merging of new code, the application is containerized using Docker and provided to ECS to be run as an automated task in the workflow.

\section{frontend UI}
\begin{figure}
    \centering
    \includegraphics[width=\linewidth]{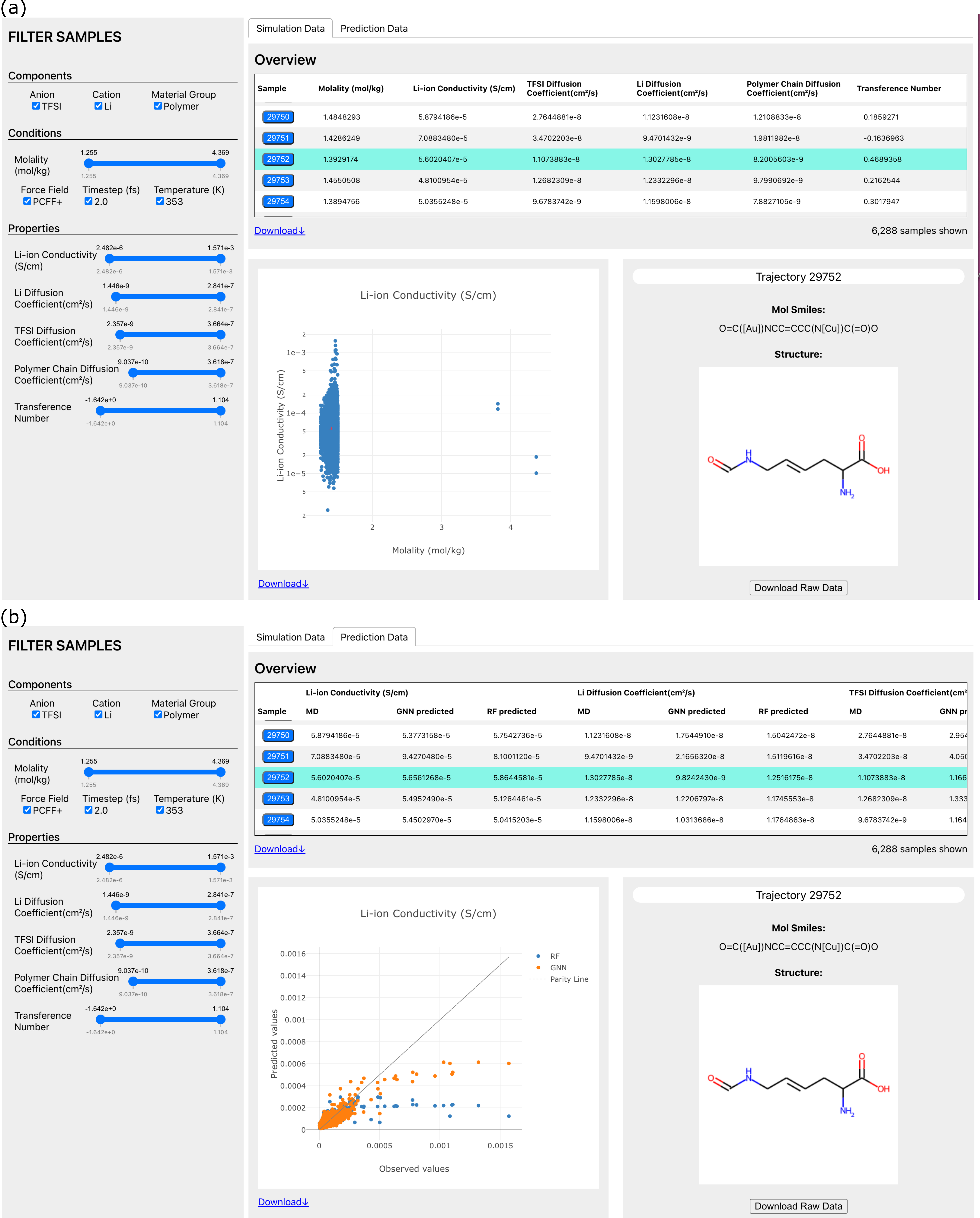}
    \caption{Web graphical user interface (GUI) to browse database for Simulation Data (a) and Prediction Data (b).}
    \label{fig:web-ui}
\end{figure}
\begin{figure}
    \centering
    \includegraphics[width=\linewidth]{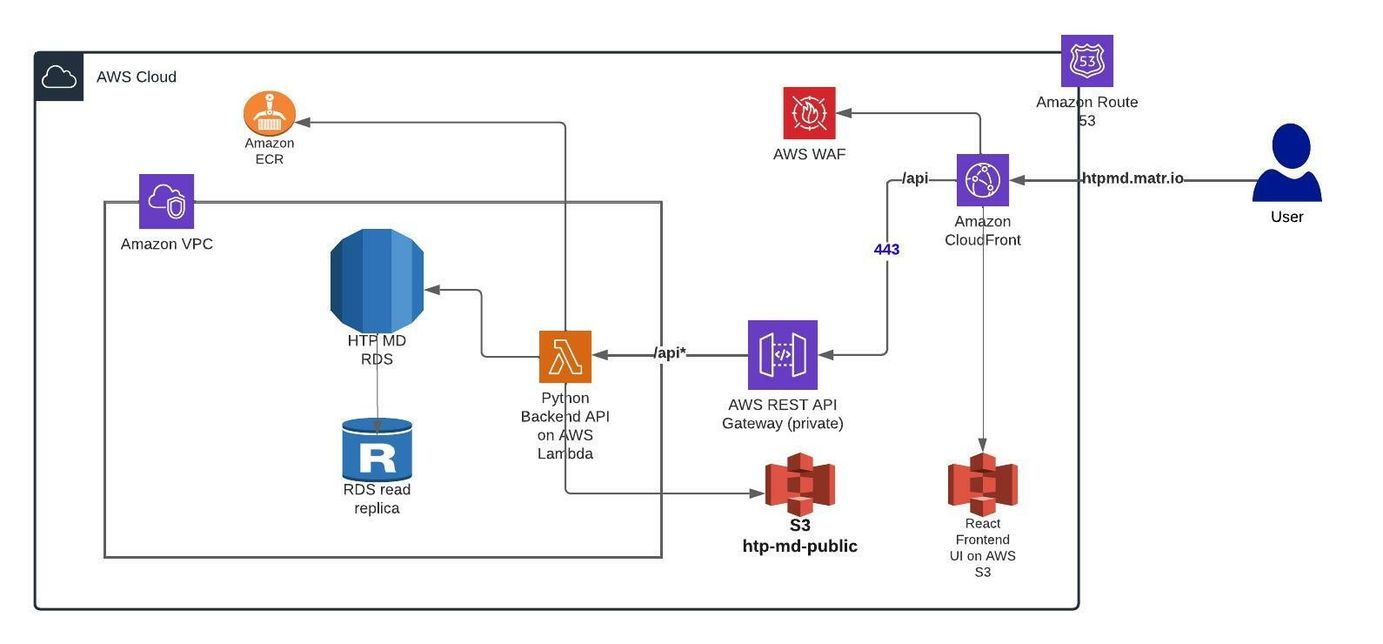}
    \caption{Workflow for retrieving data from the backend and displaying it on the frontend}
    \label{fig:ui-workflow}
\end{figure}
The platform provides a publicly accessible graphical user interface (GUI) that allows browsing through the available data. This web app is hosted at~\url{www.htpmd.matr.io} and utilizes React framework~\cite{React} for the frontend and Python for the backend. The frontend communicates with the backend using RESTful API calls, as described in Figure \ref{fig:ui-workflow}. Plots are drawn using Plotly JS, which allows graphs to be zoomed in and exported to png.

On load of the application, the frontend makes an API call to fetch all trajectories available. This is used to populate the trajectories table and generate scatter plots of properties of interest. When filters are changed on the left panel, the cached data is filtered based on the user’s selected filters.

\subsection{Simulation Data}
Frontend UI displays data in two available tabs. The default tab, Simulation Data, displays all data related to the MD simulations of polymers, such as extracted properties and simulation input parameters for individual trajectories. Properties, such as Li-ion conductivity, diffusion coefficients of Li$^+$, TFSI$^-$, and polymer chains, and the transference number, are extracted using analysis functions in the htpmd github repository. 

This tab shows aggregate data in two ways: 1) scatter plots of Li-ion conductivity and diffusion coefficients of Li ion, TFSI ion, and polymer chains as a function of composition and 2) a table overview of all trajectories (named Sample) and their properties (molality and transference number, in addition to all plotted properties). By default, analyzed properties of all available MD trajectories are shown in the graphs and tables. A filter on the side bar allows the user to down-select data by material group, cation/anion types, as well as temperature and molality ranges. In addition, the user can filter by the range of the property of interest. 

When a trajectory is selected via table row click or a plot data point click, additional API calls are made to fetch data specific to a single trajectory. These calls fetch monomer SMILES string, chemical structure, the conditions of simulations, and the Li$^+$ and TFSI$^-$ MSD time series. If the user clicks on the “Download Raw Data” button, a pre-signed S3 url is opened for the user to download a zip file of the trajectory’s raw data.

Both table and graphs have a download button which allows the user to retrieve the displayed data as comma-separated values (csv) data files. 

\subsection{Prediction Data}
Switching to the Prediction Data tab allows the user to view aggregate and trajectory-specific prediction data made using pre-loaded ML models, as shown in Figure \ref{fig:web-ui}b. The table shows Li-ion conductivity, diffusion coefficients of ions and polymers, and transference number that have been extracted from MD simulations, compared against predictions using RF and GNN models (details provided in a subsequent section). 

Scattered plots show parity plots for conductivity and ion diffusion coefficients for predicted data against MD simulation data. The user can select a specific trajectory by clicking on a data point in the plots or by selecting a row in the tables. As with the Simulation Data tab, any aggregate data can be downloaded using the download button below the table or graphs. 

\section{Polymer Database and content}

\subsection{Overview of the polymer database}

\begin{figure}[tbh]
    \centering
    \includegraphics[width=\linewidth]{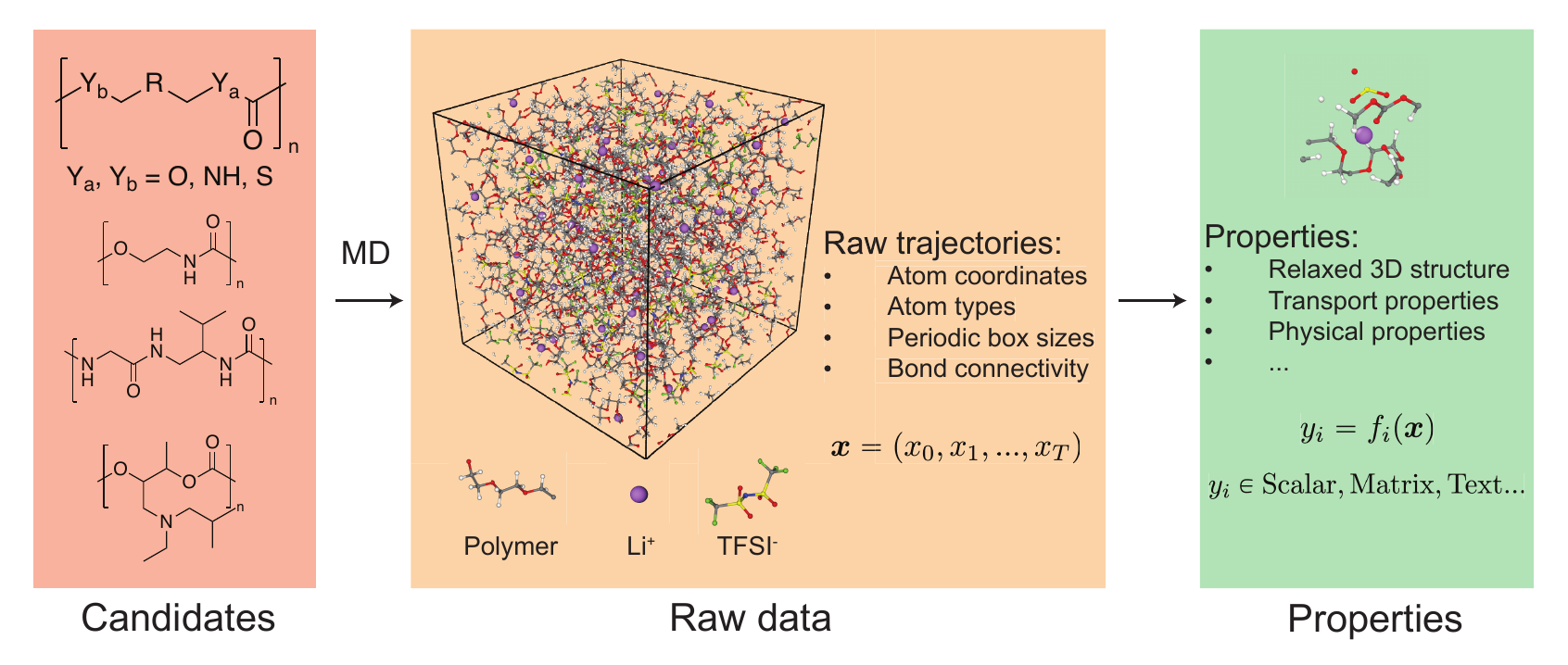}
    \caption{Database of polymer electrolytes, containing 6057 unique polymers and their raw trajectories. Raw trajectories contain coordinates and types of atoms, as well as periodic box sizes and bond connectivity. Using htpmd github repository, multiple properties including relaxed 3D structures, transport properties, and physical properties are automatically extracted from the raw trajectories.}
    \label{fig:polymer}
\end{figure}

As a demonstration of the platform, we upload the raw trajectories generated by a previous study \cite{xie2021accelerating} that uses molecular dynamics (MD) to screen polymer electrolytes for Li-ion battery applications. The database contains 6057 unique polymers that share the same structure template in \autoref{fig:polymer}, which can be synthesized through a condensation polymerization route detailed in \cite{xie2021accelerating}. The initial 3D structures of the polymer electrolytes are generated by inserting 1.5 mol lithium bis(trifluoromethanesulfonyl)imide (LiTFSI) salt per kilogram of polymer into a mixture of polymer chains and performing a 5 ns MD equilibration at 353 K. Currently, the database contains 6152 MD trajectories for 5 ns simulations of polymer electrolytes at 353 K, recorded every 2 ps. This database is more comprehensive than the previous study \cite{xie2021accelerating}, in which only 900 of these polymers were simulated with MD, the rest screened with ML property predictors. In addition, the database also contains 134 MD trajectories for 50 ns simulations of polymers, recorded every 2 ps, which provides better converged transport properties like diffusion coefficients and Li-ion conductivity. 

\subsection{Properties computed and associated methods}

Several properties are computed by default, mostly related to ion transport. Current methods rely on the identification of ion clusters, as shown in Figure \ref{fig:polymer} to calculate transport properties. An extensive list is given in Table~\ref{tab_properties}, with the associated methods. 

\begin{table*}[h]
\caption{\label{tab_properties}
Properties computed and associated methods.
}
\begin{tabularx}{\textwidth}{cccX}
\hline
 Property & Symbol and units & Type & Method and comments \\
\hline
Molality & $m$ $[\text{mol}/\text{kg}]$ & scalar & Number of ion pairs divided by the total polymer mass. \\
Structure & & string & Written out in the CIF file format~\cite{brown_cif}\\
Atomic displacement & $\delta(\left|\mathbf{X}\right|)$ $[\text{\AA}]$ & scalar & Can output either the mean or the maximum displacement along the trajectory. \\
Mean squared displacement & $\text{MSD}(t)$ $[\text{\AA}^2]$ & vector & $\text{MSD}(t) = \displaystyle\left\langle \left| \mathbf{X}(t) - \mathbf{X}(0) \right|^2 \right\rangle$ The average is performed on all atoms of the same species, and can be switched on for time origins. \\
Ion diffusivity & $D$ $[\text{cm}^2/\text{s}]$ & scalar & $D = \displaystyle\frac{1}{6} \frac{\text{d}\text{MSD}(t)}{\text{d}t}$\\
Polymer diffusivity & $D$ $[\text{cm}^2/\text{s}]$ & scalar & Defined as the average of electronegative sites (N, S, O). \\
Ionic conductivity & $\sigma$ $[\text{S}/\text{cm}]$ & scalar & Cluster Nernst-Einstein approximation~\cite{PhysRevLett.122.136001}\\
Cation transference number & $t^+$ & scalar & Cluster Nernst-Einstein approximation~\cite{PhysRevLett.122.136001}\\
\hline
\end{tabularx}
\end{table*}

\subsection{ML predictions}

The dataset enables investigating through machine learning  the transport properties of polymer electrolytes. In this work, we use two baseline machine learning models to learn the transport properties: 1) human-engineered descriptors + random forest model and 2) graph neural networks (GNNs). Twenty percent of the data points are randomly reserved at the very beginning as the test set, and the remaining data points are used to train and tune the hyper-parameters via a four-fold cross-validation. The human-engineered descriptors are generated using the package Mordred~\cite{moriwaki2018mordred}. The random forest models are built using scikit-learn~\cite{pedregosa2011scikit}. We adopt the GNN architecture used in our prior work \cite{xie2021accelerating} that builds Crystal Graph Convolutional Neural Networks (CGCNN) \cite{xie2018crystal} on top of polymer graphs.  Note that in this work, both models are based on 2D molecular information of the monomer; machine learning models making use of the 3D structure of polymers will be discussed in the future.

In Table 2, we show the performance of the machine learning models on five transport properties: Li-ion conductivity, Li$^+$, TFSI$^-$, and polymer chain diffusion coefficients, and transference number. We can see that the deep representation learning model (GNN) performs slightly better than the random forest model based on human-engineered descriptors for all transport properties, except for the  transference number. Also, except for the polymer chain diffusion coefficient, the $R^{2}$ scores of prediction of transport properties from both machine learning models are lower than 0.8, which indicates the limitations of current two models. Model performance can be potentially further improved by including 3D features or even dynamical features. 

\begin{table*}[h]
\setlength{\tabcolsep}{16pt}
\caption{\label{tab_hyperpara}
Performance of different machine learning models on Li transport properties. Errors are based on the test set, and error bars are standard deviations of predictions from the four models trained in the four-fold cross-validation.
}
\begin{tabular}{l|cccc}
\hline
Property & \multicolumn{2}{c}{RF + molecular features} & \multicolumn{2}{c}{GNN + molecule structures}\\
\hline
 & MAE & R2 & MAE & R2 \\
$\sigma$ [S/cm] & $0.120 \pm 0.003$ & $0.532 \pm 0.024$ & $0.115 \pm 0.002$ & $0.573 \pm 0.014$\\
$D_{\text{Li}^+}$ [cm$^2$/s] & $0.117 \pm 0.002$ & $0.508 \pm 0.012$ & $0.115 \pm 0.000$ & $0.492 \pm 0.005$\\
$D_{\text{TFSI}^-}$ [cm$^2$/s] & $0.103 \pm 0.002$	& $0.633 \pm 0.017$ & $0.100 \pm 0.001$ & $0.650 \pm 0.010 $\\
$D_{\text{chain}}$ [cm$^2$/s] & $0.088 \pm 0.002$	& $0.820 \pm 0.005$ & $0.082 \pm 0.001$ & $0.832 \pm 0.002 $\\
$t^+$ & $0.158 \pm 0.000$	& $0.502 \pm 0.002 $ & $0.159 \pm 0.001$ & $0.491 \pm 0.004 $\\
\hline
\end{tabular}
\end{table*}

\section{User Scenario}
We envision two broad user scenarios for the platform: 1) A user whose primary goal is to explore and visualize existing data (raw data and analyzed properties); 2) A user who wishes to develop and contribute new analysis functions and ML prediction models to derive additional insights from existing data, or use existing analysis functions and ML prediction models on private data and potentially contribute new data to our platform. We outline recommended workflows for each user scenario. 

\subsection{Visualization and exploration of data}
A user wishing to explore the data can access the platform at htpmd.matr.io. All trajectory data are loaded at once; however, the user can down-select the data using the selection panel on the left, filtering by components, simulation conditions (molality, force field, time step, simulation length, and temperature), and the desired range of analyzed properties. Most trajectories also have additional data (chemical structure and mean squared displacement time series) and can be filtered by whether the data are available.  

\begin{figure}[tbh]
    \centering
    \includegraphics[width=0.8\linewidth]{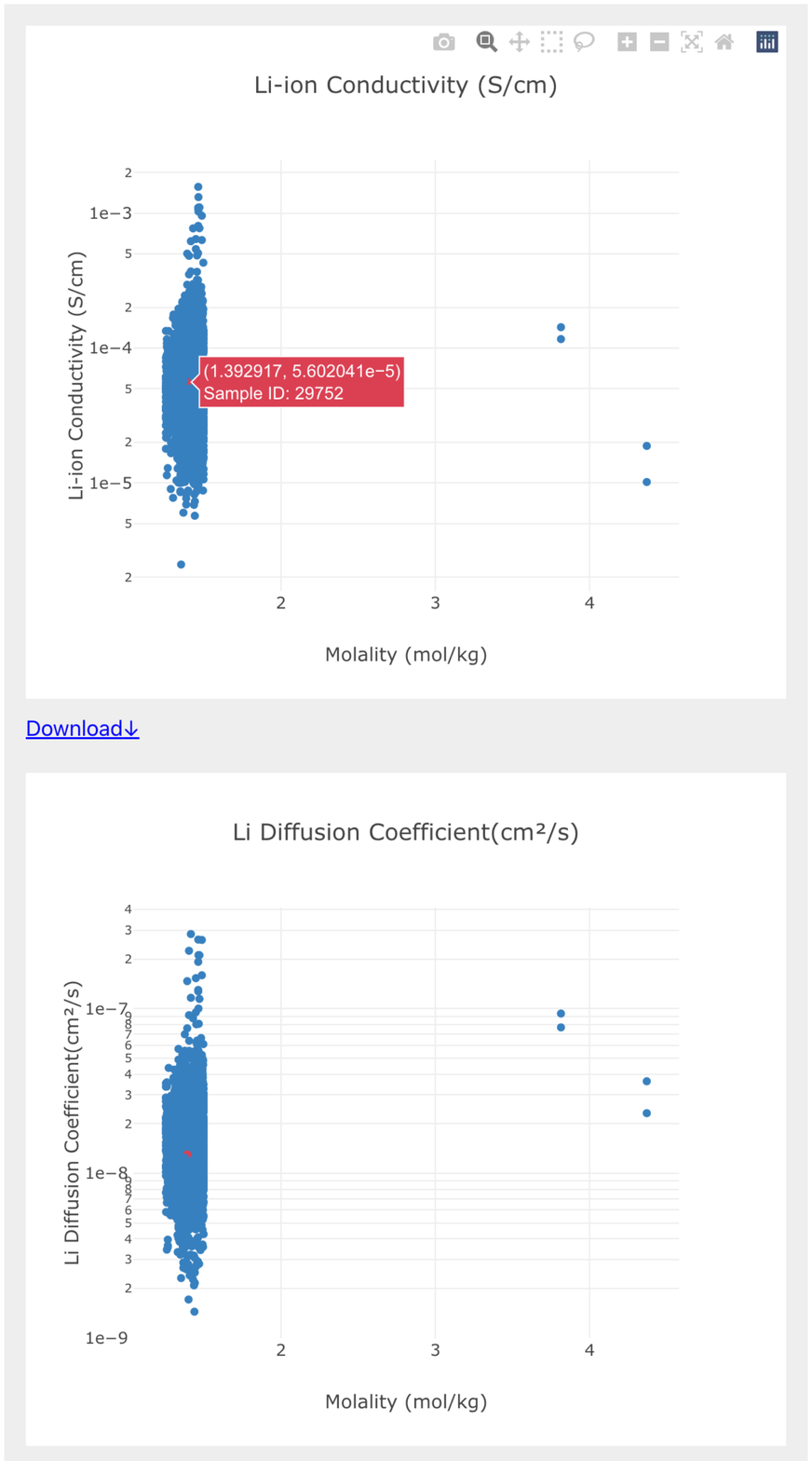}
    \caption{Aggregate data plots showing Li-ion conductivity and Li$^+$ diffusion coefficient of all available trajectory data. Numeric values and trajectory ID can be shown when hovering on a specific data point. Clicking on a data point highlights all corresponding points in subsequent graphs, as well as on the table. This also enables trajectory sample view, which gives specific information about the polymer, simulation conditions, and property values.}
    \label{fig:aggregate}
\end{figure}

Filtered data are displayed in tabular and graphical formats. The data table lists simulation conditions and analyzed properties for each trajectory ID and can be sorted by ascending/descending value. Aggregate view shows plots of three relevant properties (Li-ion conductivity and diffusion coefficients of Li$^+$ and TFSI$^-$) as a function of molality (Figure \ref{fig:aggregate}). Users can hover over each data point for trajectory ID information or zoom into parts of each plot using click-and-drag. 

More detailed information on a single trajectory can be displayed by selecting a specific trajectory (in the table or on the graph), as shown in Figure \ref{fig:web-ui}. The sample view displays MSD time series, chemical structure, and simulation conditions for the selected trajectory.

Trajectory-specific data in sample view or aggregate information in aggregate view can be downloaded by clicking on the Download button. This information is downloaded as a csv file. 

\subsection{Community contribution of new analysis methods and data}
Some users may wish to run the analysis module locally in order to develop new analysis functions, train new ML prediction models, or try existing analysis functions on private data. This can enable new insights from existing data. For these users, the best starting point is at the public github repository at \url{https://github.com/tri-amdd/htp_md/src}. Users are encouraged to install within a Docker container to minimize dependency issues. 

\begin{figure}[tbh]
    \centering
    \includegraphics[width=.5\linewidth]{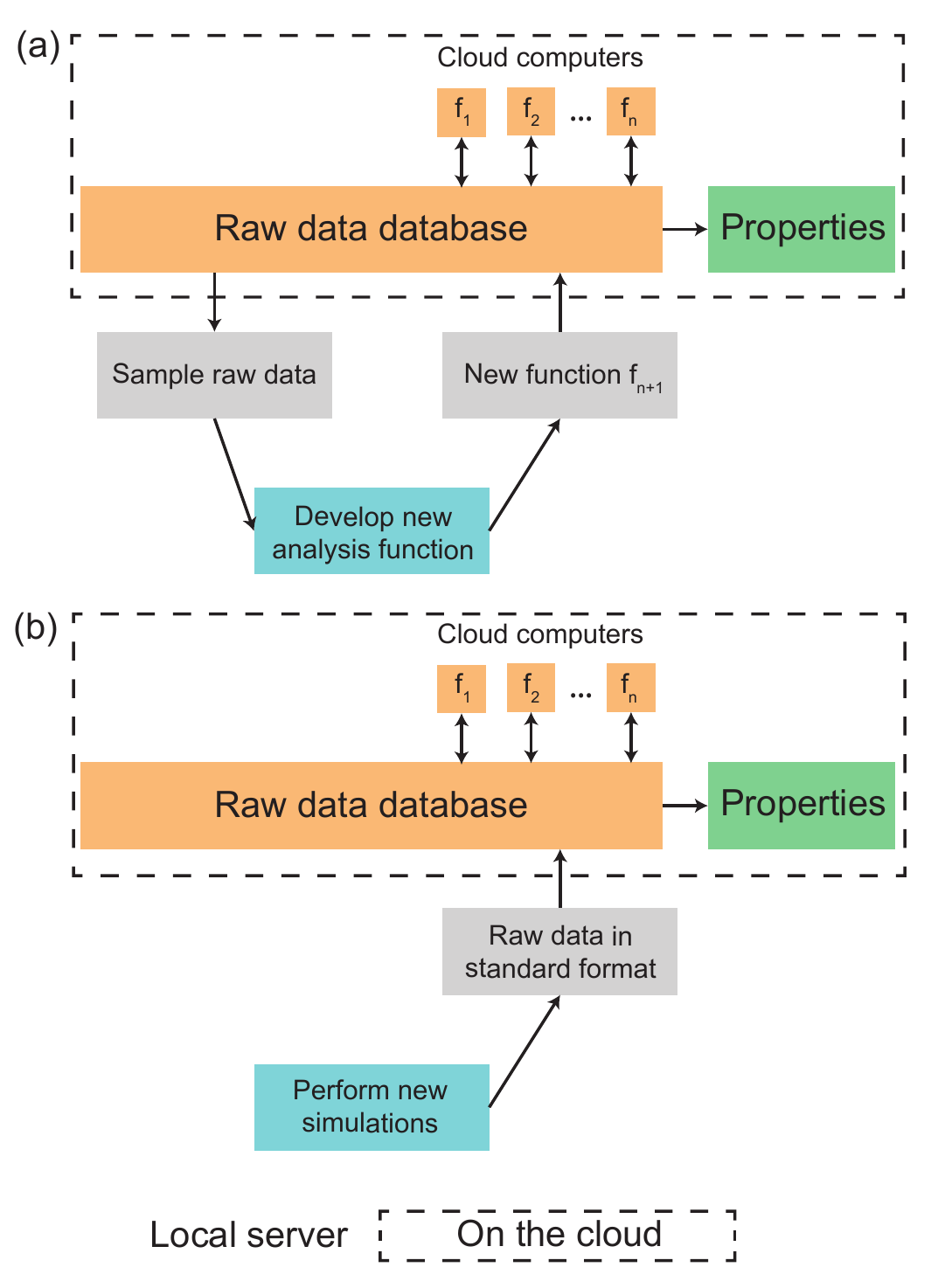}
    \caption{Schematics for contributing new methods and data. (a) Analysis method contribution. User can download sample raw data from the database and develop new analysis function locally. New functions will be scaled to the entire database after merging. (b) Data contribution. User can contribute new trajectories to the database by following the standard format. All existing analysis functions will be used to analyze the new data.}
    \label{fig:user_scenario}
\end{figure}

The repository provides test data for three test systems (an aqueous NaCl electrolyte and two LiTFSI polymer electrolytes). Additional data can be downloaded from the database via the UI. Newly developed analysis functions can be tested on the test data, as well as any data downloaded from the database (see Figure \ref{fig:user_scenario}a)

In order to merge the new analysis function or method, the user must open a pull request containing the source code in \url{function.py} and an accompanying test in \url{function_test.py}, as well as test data and results (if different from provided test data). The format of the code should follow the provided template. Contributed code will be reviewed by the repository maintenance team. 

Once contributed code is reviewed and merged, htpmd version number will be updated, and the pipeline will run the latest version of analysis functions on existing data. Latest versions of extracted properties will be available on the UI.

Alternatively, some users may wish to contribute new data that were locally generated (Figure \ref{fig:user_scenario}b). 
Users are not required but are encouraged to provide information on the compiler and version of LAMMPS used. 

\section{Discussion}

The htpmd database enables researchers to harvest insights from molecular dynamics simulations of thousands of polymer-salt systems. We will be adding trajectories and property data as we simulate more polymer-salt systems to identify their respective ion transport properties. We encourage researchers in the community to make use of the presented data, methods, and models for their own investigations and for use as benchmarks. We also welcome any contributions to this database. If you would like to add your data, methods, or machine learning models to this platform, please contact us for details. In future updates of the database web portal, we envision adding functionality that will enable direct upload and automatic verification of new data from MD simulations, experiments, and literature.

In order to make material simulations a meaningful tool in the pursuit of accelerated materials discovery, it is necessary to establish the validity and accuracy of material property data resulting from such simulations. Previous work examined the alignment of MD simulations results with experimentally determined transport properties, such as Li$^+$ conductivity~\cite{doi:10.1021/jp3052246, doi:10.1021/acs.jpcb.7b08258, doi:10.1021/acs.jpcb.0c05108}. To further complete the picture, we envision future additional features of the platform which allow a direct comparison of simulation data to experimental results published in the literature. This will enable researchers to further explore the conditions for validity, possible limitations, and future improvements to the simulation methodologies. 

A separate effort which is currently underway at MIT in collaboration with TRI is the development of high throughput experimental screening methods for measuring ion conductivity in solid polymer systems. One of its uses will be to validate computational results via synthesis and characterization of previously simulated systems. The experimental data generated through the screening can be further incorporated into our database. 

Finally, the cloud-based platform and its infrastructure can be further extended into other types of simulation approaches and material properties. The current available analysis functions and the frontend UI are specific to the transport properties in polymer-salt systems. However, the backend workflow only requires a raw data format and a list of analysis functions designed for the format. It means that we can extend the platform to new types of data by creating a standard raw data format and corresponding analysis functions. Our platform is most suitable for use cases where a large number of useful properties can be extracted from the same raw data, and the computational cost of postprocessing is significantly cheaper than the simulation itself. As the use of this platform increases, we envision broadening the scope of our platform to new problems like mechanical and rheological properties of complex polymer-salt systems.

\section{Data availability}

All data are available at \url{https://www.htpmd.matr.io}.

\section{Code availability}

Code is available at \url{https://github.com/TRI-AMDD/htp_md}.

\section{Acknowledgement}

This work was supported by Toyota Research Institute. Computational support was provided by the National Energy Research Scientific Computing Center, a DOE Office of Science User Facility supported by the Office of Science of the U.S. Department of Energy under Contract No. DE-AC02-05CH11231, and the Extreme Science and Engineering Discovery Environment, supported by National Science Foundation grant number ACI-1053575.

\section{Author contributions}

T.X., H.K.K., D.S., J.C.G., Y.S.H. conceived the idea. H.K.K., D.S., T.X. led the development of the platform. T.X., S.G., A.F.L., E.C. generated the data and developed the analysis functions. H.K.K., D.S., A.K., M.P., C.F., W.P. developed the software infrastructure on AWS, including both frontend and backend. All authors (T.X., H.K.K., D.S., S.G., A.F.L., A.K., E.C., M.P., C.F., W.P., Y.S.H., J.C.G.) contributed to the writing of the paper.

\bibliography{ref}

\end{document}